\documentclass{article}
\usepackage[utf8]{inputenc}
\usepackage{graphicx}
\usepackage{amsmath,calc, mathtools, url, cleveref, graphicx}
\usepackage{amssymb}
\usepackage[euler]{textgreek}
\usepackage{subfigure}
\usepackage{makeidx}
\usepackage{multicol}
\usepackage{bm}
\usepackage{gastex}
\usepackage{natbib}
\usepackage[dvipsnames]{xcolor}

\title{Eliciting judgements about dependent quantities of interest: The SHELF extension and copula methods illustrated using an asthma case study}

\author{Bj\"{o}rn Holzhauer$^1$ \and Lisa V. Hampson$^1$ \and John Paul Gosling$^2$ \and Bj\"{o}rn Bornkamp$^1$ \and Joseph Kahn$^3$ \and Markus R. Lange$^1$ \and Wen-Lin Luo$^3$ \and Caterina Brindicci$^1$ \and David Lawrence$^1$ \and Steffen Ballerstedt$^1$ \and Anthony O'Hagan$^4$}

\date{%
    $^1$Novartis Pharma AG, Analytics, Basel, Switzerland\\%
    $^2$JBA Risk Management Ltd, Skipton, United Kingdom\\%
    $^3$Novartis Pharmaceuticals Corporation, Analytics, East Hanover, NJ, USA\\%
    $^4$The University of Sheffield, School of Mathematics and Statistics, Sheffield, United Kingdom\\[2ex]%
    \today
}

\begin{document}
\maketitle 

\begin{abstract}
	Pharmaceutical companies regularly need to make decisions about drug development programs based on the limited knowledge from early stage clinical trials. In this situation, eliciting the judgements of experts is an attractive approach for synthesising evidence on the unknown quantities of interest. When calculating the probability of success for a drug development program, multiple quantities of interest --- such as the effect of a drug on different endpoints --- should not be treated as unrelated.
	
	We discuss two approaches for establishing a multivariate distribution for several related quantities within the SHeffield ELicitation Framework (SHELF). The first approach elicits experts' judgements about a quantity of interest conditional on knowledge about another one. For the second approach, we first elicit marginal distributions for each quantity of interest. Then, for each pair of quantities, we elicit the concordance probability that both lie on the same side of their respective elicited medians. This allows us to specify a copula to obtain the joint distribution of the quantities of interest.
	
	We show how these approaches were used in an elicitation workshop that was performed to assess the probability of success of the registrational program of an asthma drug. The judgements of the experts, which were obtained prior to completion of the pivotal studies, were well aligned with the final trial results.
\end{abstract}

\section{Introduction}

The decision to continue or stop the development of a new drug is an example of high-stakes decision making in the pharmaceutical industry. To continue usually means a commitment to large and costly clinical trials that may expose the enrolled patients to risks, while to stop may mean a missed opportunity to help patients. At the same time, only limited data are usually available. Thus, improving the decision making in these situations is an important problem.

For decision making with no or limited directly relevant data, eliciting the judgements of a group of experts is one approach to effectively combining the available direct and indirect evidence. Expert knowledge elicitation is the process of capturing expert knowledge about one or more uncertain quantities in the form of a probability distribution.  It is an important tool to provide understanding of uncertain phenomena and inputs to decision-making processes.  There has been a steadily growing demand for elicitation in many fields throughout industry, government and science --- see, for example,~\citet{Garthwaite2000}, \citet{Gosling2012}, \citet{Usher2013}, and \citet{Bamber2019}. In particular, elicitation has been advocated and used in pharmaceutical science \citep{Kinnersley2013,Dallow2018} and public health \citep{Ren2017,Soares2018}. Due to the cognitive biases that experts are subject to, several frameworks and procedures have been proposed to guide the elicitation process in order to minimise these biases. The SHeffield ELicitation Framework (SHELF) described in Section~\ref{sec:shelf} of this paper is one such framework.

It can be challenging to elicit judgements about a quantity of interest (QoI), when these judgements are being made conditional on knowledge about another quantity. Similarly, QoIs are often likely to be dependent, in which case the challenge of eliciting a joint distribution for several QoIs arises. There are many methods in the literature for capturing knowledge about dependencies between multiple variables~\citep{Daneshkhah2010,Werner2018}. However, these methodologies are typically reported in the literature as standalone methods rather than forming part of a complete elicitation protocol like SHELF. Also, whereas SHELF is a generic protocol that is applicable to a very wide range of applications, most of these methodologies have considerable restrictions.

\begin{itemize}
    \item They may constrain the type of variables and distributions to be fitted --- for example, to Dirichlet distributions for proportions~\citep{Elfadaly2013, zapata2014eliciting}.
    \item They may be tailored for a specific application --- for example, land cover~\citep{Baey2017} or system reliability~\citep{Norrington2008}.
    \item They may consider complex restructuring for large numbers of dependent variables \citep{Sigurdsson2001,Bedford2010,Truong2013}.
\end{itemize}  

We present generic methods for eliciting joint distributions through judgements that experts can realistically make. Like the SHELF protocol itself, these methods are applicable in all areas where elicitation is required, and to use them effectively there are important choices to be made. Examples of its use, and the choices made, in any specific field can therefore serve as valuable guides for others to follow.  We illustrate their use within SHELF in a pharmaceutical example that we fully describe in Section~\ref{sec:motivate}. The example concerns assessing the probability of success (PoS) of a Phase 3 drug development program. Such programs are expensive, resource-intensive long-term commitments for any organisation. 
The decision to proceed with a Phase 3 program depends on many considerations including the unmet medical need and market opportunity a new drug may address, as well as the probability of success to address these needs.
As part of a pilot project to evaluate a new PoS framework at Novartis, we conducted a PoS assessment for an asthma drug. While Phase 2 studies had provided information on the effect of the drug on a surrogate outcome, no data were available on the primary endpoint of the key Phase 3 studies: moderate-to-severe asthma exacerbations, which are potentially life-threatening events with a significant burden on patients' lives~\citep{GINA2020}. Additionally, there was an important key secondary endpoint --- forced expiratory volume in 1 second (FEV\textsubscript{1}), an endpoint commonly used in asthma trials ---, for which Phase 2 data were available, but experts' judgements were sought on the effect of the different  treatment duration and trial population in Phase 3. If the drug worked on one endpoint, it was considered to more likely work on the other endpoint. Thus, a joint distribution was required. Techniques to address both problems through expert elicitation are available within the SHELF framework.

In Section~\ref{sec:shelf}, we first give a brief overview of elicitation methods and of SHELF. Then we describe the extension method for eliciting judgements about Phase 3 outcomes by linking to Phase 2 results and the copula method for eliciting joint distributions. In Section~\ref{sec:qaw039}, we return to our motivating example and describe how we used these techniques to estimate the PoS of that drug development program. We also compare the obtained expert judgements with the outcomes of the Phase 3 studies. We finish with some conclusions and recommendations in Section~\ref{sec:discussion}.

\section{Motivating example} \label{sec:motivate} 

The Phase 3 program of fevipiprant, a prostaglandin D\textsubscript{2} receptor 2 antagonist for the treatment of asthma, was selected to pilot a new PoS framework that has since been introduced at Novartis~\citep{hampson2021}. At the time of the PoS assessment, fevipiprant had been studied in several Phase 2 randomised controlled trials (RCTs) and the Phase 3 clinical trials comparing two fevipiprant doses (150 or 450 mg once a day) with placebo were underway, with data collection almost complete. This timing was one reason the program was selected as a pilot, because it ensured that the PoS assessment could not be influenced by the Phase 3 data, while at the same time minimising the time until the PoS assessment could be compared to the Phase 3 results. In reality, the assessment of the program and the decision to proceed with Phase 3 had already been taken at the end of Phase 2 based on more limited information.

One major challenge was that --- unlike the Phase 2 trials --- the key Phase 3 trials focused on more severe asthma patients with the sub-population with a blood eosinophil count $\geq$ 250 cells/\textmugreek l. The primary null hypotheses for this sub-population were tested first in the trials' testing procedures~\citep{Brightling_2020}. None of the Phase 2 trials evaluated the effect of fevipiprant on moderate-to-severe asthma exacerbations. The annualised rate of such exacerbations was the primary endpoint of the two most important trials in the Phase 3 program~\citep{Brightling_2020}. Instead, a surrogate endpoint of reduction in sputum eosinophil counts had been measured in one of the Phase 2 trials~\citep{Gonem_2016}. FEV\textsubscript{1} was a key secondary endpoint in the Phase 3 program and has high regulatory acceptance as a measure of asthma control~\citep{ema2015asthma}. FEV\textsubscript{1} had been a primary or secondary endpoint of several of the Phase 2 studies including for dose ranging~\citep{Bateman_2017}, but these trials were of shorter duration and had a patient population with milder asthma than the Phase 3 trials.

As per the newly implemented PoS framework at Novartis, success was defined as regulatory approval with point estimates for key endpoints achieving or exceeding targets specified as part of a target product profile (TPP). It was assumed that regulatory approval would require statistical significance at the one-sided 0.025 significance level for at least one dose for both exacerbations and FEV\textsubscript{1} in both of the key Phase 3 trials. Thus, to calculate the PoS, we needed a joint prior distribution for the effects of fevipirant on exacerbations and FEV\textsubscript{1}.

Given the data that were available at the time of the PoS assessment, we decided to do this by eliciting the judgements of a group of experts. The question then was how best to structure the elicitation process: we wanted to explicitly leverage the Phase 2 data on the surrogate endpoint of reduction in sputum eosinophil counts, since this was arguably the most relevant evidence we had for informing beliefs about the effect of fevipiprant on exacerbations. 

We also expected experts to judge a larger effect of fevipiprant on FEV\textsubscript{1} to be more likely the larger the effect of the drug on asthma exacerbations is. As a consequence, in order to fully characterise the joint distribution of these two treatment effects we would need to understand the size and direction of the dependence between these two quantities. In the next section, we describe the various approaches considered for the elicitation, before we return to the motivating example in Section~\ref{sec:qaw039} and describe how we practically applied these methods.

\section{Overview of SHELF}\label{sec:shelf}

\subsection{Elicitation protocols}
\label{subsec:protocols}

Elicitation can be done informally, but numerous pitfalls await the inexperienced practitioner, including well-established sources of bias in expert judgements \citep{Ohagan2006,EFSA2014,Ohagan2019}. Therefore, when the expert judgements are sufficiently important it is necessary to employ a formal procedure in the interests of quality and defensibility. A small number of established elicitation {\em protocols} have been developed and refined by experienced practitioners \cite[an overview is given in][]{Dias2018}.

The SHELF protocol is characterised by carefully structured sequences of judgements designed to minimise biases and a unique way of eliciting a consensus probability distribution from a group of experts \citep{Gosling2018}. It is one of the most widely used elicitation protocol in the field of pharmaceutical science. The SHELF package of advice, templates and tools to support researchers wishing to conduct expert knowledge elicitation may be freely downloaded from the SHELF website~\citep{shelfwebpage2019}. Since its inception in 2008, SHELF has been steadily expanded with new advice and methods. For example, the extension method described in Section~\ref{subsec:shelf extension} was added in version 4~\citep{shelf2019shelfv4}. 

\subsection{The basic SHELF method and principles}
\label{subsec:shelf basics}

The SHELF protocol is distinguished by a number of key elements.
\begin{itemize}

    \item
Individual elicitation –- discussion –- group elicitation.  Serious elicitation almost always requires using a group of experts in order to capture their combined knowledge.  SHELF elicits a single distribution from the group but begins by eliciting judgements from each expert independently.  This is followed by the experts discussing their differences to share their expertise, opinions and interpretations of the evidence. Finally, group judgements are elicited and the result is a “consensus” distribution fitted to these judgements.  This combination of individual and group elicitations is the most important distinguishing feature of SHELF. The individual elicitations show each expert's beliefs and form a basis for the subsequent discussion. The discussion is an opportunity to share and debate those opinions with a view to achieving a common understanding and is intended to extract maximum value from their joint expertise,

\item
The SHELF workshop.  The discussion and group elicitation phases require that the experts come together in what is called a SHELF workshop.  Typically, they are physically together in a room, although SHELF can be used with other arrangements, including video-conferencing.

\item
The evidence dossier. Prior to the workshop, a dossier is prepared summarising the evidence regarding the QoIs. In a typical elicitation there is some relevant evidence available, but there is not enough direct evidence to identify the value of any QoI (otherwise expert judgement would not be needed). The dossier ensures that all experts have access to the same evidence and that it is all fresh in their minds when they make their judgements. The essence of expert knowledge elicitation is that different experts interpret and weight the evidence differently, based on their own experience. The discussion phase in SHELF is where these differences are aired and debated.

\item
The rational impartial observer (RIO).  Even after discussing and debating, SHELF does not expect the experts to reach complete agreement (such that they now have the same knowledge and beliefs about an uncertain quantity, represented by the same probability distribution).  Instead they are asked to judge what a rational impartial observer, called RIO, might reasonably believe, having seen their individual judgements and listened to their discussion.  By taking the perspective of RIO, experts can reach agreement on a distribution that represents a rational impartial view of their combined knowledge.

\item	
The facilitator.  The SHELF workshop is led by a facilitator, who has expertise in the process of eliciting expert knowledge, and in particular is familiar with SHELF. The facilitator works with the experts to accurately capture their knowledge, facilitates the group discussion and leads them in applying the RIO perspective.  The facilitator’s role may also be found in other elicitation protocols, but it is particularly important in SHELF. The group interaction in SHELF's discussion is another possible source of biases, which must be managed by the skill and experience of the facilitator.  Many other protocols do not admit group discussion, thereby avoiding the risk of those biases but also losing the opportunity for the experts to share and debate their judgements.

\item
SHELF templates.  The conduct of the workshop is recorded on SHELF templates, which play a dual role.  First, they organise the progress of the workshop through a predefined series of steps.  In particular, both the individual elicitations and the group elicitation are directed through controlled sequences of judgements.  The entire process and the elicitation sequences used are based on research into the psychology of judgement, and on extensive experience in practical elicitation. Second, the templates serve to document a SHELF workshop, such that conduct of the workshop and the development of each elicited distribution is clearly set out.
\end{itemize}

The result of an elicitation for a single uncertain QoI is a probability distribution. Accordingly, the judgements that experts are asked to make are probabilistic. The basic sequence of judgements at the individual judgements stage is as follows.

\begin{enumerate}

\item
Plausible range. Experts are first asked to specify upper and lower plausible bounds such that they judge values of the QoI outside that range to be implausible. A numerical interpretation of `implausible', for instance as a 1\% or 5\% probability, is not generally made, since the primary function of this step is to encourage the experts to think of all possibilities, thereby reducing any tendency to overconfidence.
\item
Median. Experts are next asked to specify their median value for the QoI, such that they regard it as equally likely that the QoI would be above or below this value.
\item
Quartiles or tertiles. Finally, experts specify their quartile or tertile values (the choice of which to ask for being according to the facilitator's preference). Just as the median divides the plausible range into two intervals that are judged equally likely, quartiles divide it into four equally likely intervals and tertiles into three. In expert elicitations for the Novartis PoS framework we have favoured eliciting tertiles instead of quartiles, because we consider thinking about three instead of four equally likely intervals less challenging for experts.

\end{enumerate}

The SHELF package contains copious advice and tools to help the experts to understand and make these judgements reliably. In particular, by following the SHELF protocol the facilitator asks questions in such a way that biases are minimised and there is no need for the experts to have a thorough understanding of probability or statistical theory. Training in making these judgements is also available through an online self-paced course accessed from the SHELF website~\citep{shelf_e_learning_course}.

For the group judgements, the facilitator may ask the experts to agree on probabilities that RIO might assign to three specific propositions, such as that the QoI is negative, or that it exceeds some specified value. A probability distribution is then fitted to the three RIO probability judgements. SHELF provides some R software for fitting distributions using various standard families, such as normal, t, gamma, lognormal or beta distributions~\citep{shelfRpackage2020}. However, other distributions may be fitted, and indeed a major consideration in SHELF is that the form of the elicited distribution should not be constrained in any way.  The facilitator will work with the experts to identify a suitable distribution to represent their judgements. The fitted distribution is the final outcome of the elicitation. 

Throughout the process, and particularly when determining the final agreed distribution, the facilitator will prompt and challenge the experts to ensure that the final distribution genuinely represents what RIO might believe after seeing the experts' judgements and listening to their discussions.

\subsection{The SHELF extension method}\label{subsec:shelf extension}

The SHELF package contains several techniques for eliciting a joint distribution for two or more uncertain quantities, including the extension and copula methods. The extension method is a generic technique that allows considerable flexibility for the form of the joint distribution.
It is e.g. suitable for eliciting judgements about the treatment effect for a Phase 3 endpoint ($X$) based on the Phase 2 results for a surrogate endpoint ($Y$). The fact that Phase 3 follows Phase 2 chronologically makes it natural to express judgements about $X$ conditional on $Y$.

For two QoIs, $X$ and $Y$, the extension method consists of obtaining a marginal distribution for $Y$ and a set of conditional distributions for $X$ given $Y=y$. The elicitation of joint distributions requires the following steps. 

\begin{enumerate}
    \item A marginal distribution for $Y$ is obtained. This distribution can be elicited as described in Section \ref{subsec:shelf basics}, but could also be the result of an analysis of available data. E.g. in the asthma case study introduced in Section \ref{sec:motivate} it is a meta-analytic predictive distribution~\citep{neuenschwander2010historical} based on Phase 2 data.
    
    \item A conditional distribution (as always, from the perspective of RIO) is elicited for $X$ conditional on $Y$ equalling the median of its elicited marginal distribution, also following the basic method of Section \ref{subsec:shelf basics}.
    
    \item Several other quantiles of the elicited marginal distribution of $Y$ are selected as conditioning points; typically these will be the quartiles, 5th and 95th percentiles. Median values are elicited for $X$ conditional on $Y$ equalling each of theses conditioning points (first the 5th and 95th percentiles and then the quartiles). The basic SHELF approach of individual judgements -- discussion -- group judgements is used for each.
    
    \item The final step is to `fit' a set of conditional distributions to these judgements. First, a median function $m(y)$ is fitted to the elicited conditional medians. This might for instance be a polynomial or a piecewise-linear fit (with extrapolation), and may be applied on a transformed scale. Second, a model is chosen to determine the conditional distributions based on the distribution at the $Y$-median elicited in Step 2. For instance, it may be decided that the $Y$-median distribution can be applied to all conditionals, simply shifted to follow the $m(y)$ function. Alternatively, the variance may also be scaled depending on $m(y)$. These choices are available in the SHELF R software, but again other choices can be made.  The facilitator will always work with the experts to identify a `fit' that best represents their judgements.
\end{enumerate}

The extension method is appropriate when the experts perceive a natural causal link from $Y$ to $X$. Indeed, it is particularly useful when the objective is to elicit a distribution for $X$ but the experts would find it easier to make judgements about $X$ if they knew the value of $Y$. In this case, the marginal distribution of $X$ is the main outcome of the elicitation process. Although it will not generally be feasible to derive that marginal distribution analytically from the elicited joint structure, a large Monte Carlo sample can be drawn by sampling values $y_i$ from the marginal distribution of $Y$ and then sampling $x_i$ conditional on $Y=y_i$. The Monte Carlo samples $\{x_i\}$ are then samples from the marginal distribution of $X$ and, if needed, a distribution can be fitted to the samples.

\subsection{The SHELF copula method}\label{subsec:shelf copula}

When there is no natural ordering of related QoIs based on time or causality, the extension method requires an arbitrary imposition of an ordering and the conditional judgements are more difficult for the experts.  The SHELF copula method is appropriate for two or three QoIs and does not require the elicitation of conditional distributions. However, it does place some constraints on the joint distribution.  The method has the following steps.

\begin{enumerate}
    \item Marginal distributions are elicited for each QoI individually, using the basic method of Section \ref{subsec:shelf basics}.
    \item For each pair of QoIs, a single judgement concerning their degree of correlation is made. This judgement is called the concordance probability, and is the probability that both QoIs lie on the same side of their respective elicited medians.
    \item A Gaussian copula joint distribution~\citep{trivedi2007copula} is then fitted to these marginal distributions and concordance probabilities. The facilitator shows the experts suitable displays or summaries of the joint distribution to verify that it is a reasonable representation of their beliefs.
\end{enumerate}
With just two QoIs, the copula method is simple to apply. The Gaussian copula imposes a restriction on the joint distribution but in practice it will usually be an adequate fit to the experts' judgements. 

In principle, the copula method is applicable for larger numbers of QoIs, but it is difficult to use for more than three.  With three QoIs, three concordance probabilities need to be elicited. Under the Gaussian copula assumption, each concordance probability can be transformed to a correlation coefficient and the resulting correlation matrix must be positive definite. It is quite possible for the experts' elicited concordance probabilities to fail to produce a valid correlation matrix, and they must then revisit their judgements with the aid of the facilitator to achieve an adequate fit. With more than three QoIs, the number of concordance probabilities rapidly increases, as does the likelihood of the elicited values not corresponding to a valid correlation matrix.

The SHELF copula method is a natural choice to construct a joint distribution for the effects of a drug on two Phase 3 endpoints, such as a primary and secondary clinical outcome.

The interested reader will find full technical details, as well as much practical advice, on these and other elicitation techniques in the SHELF package~\citep{shelf2019shelfv4}.
\bigskip

\section{Asthma case study} \label{sec:qaw039}

In this section, we provide an in-depth description of the expert elicitation and PoS calculation for the example introduced in Section~\ref{sec:motivate}. We decided to structure the elicitation process into three parts. First, we followed the SHELF extension method by using Phase 2 data to establish a marginal distribution for the effect of fevipiprant on sputum eosinophil counts and then elicited from a group of experts a set of conditional judgements on the effect on exacerbations in the Phase 3 population given different values for the effects on this surrogate endpoint.
Secondly, we elicited the experts' beliefs on the effect of fevipiprant on FEV\textsubscript{1} in the Phase 3 population. Finally, we used the SHELF copula method to elicit the dependence between drug effects on exacerbations and FEV\textsubscript{1}.

\subsection{Available evidence}

\begin{figure}[b!]
 \caption{Observed differences in FEV\textsubscript{1} to placebo with 95\% confidence intervals for fevipiprant in Phase 2 studies in the subgroup with a blood eosinophil count of $\geq 250$ cells/$\mu$L and in the overall trial populations (A: atopic patients, NA: non-atopic patients)} \label{fig:fev1rcts}
\includegraphics[width=0.98\textwidth]{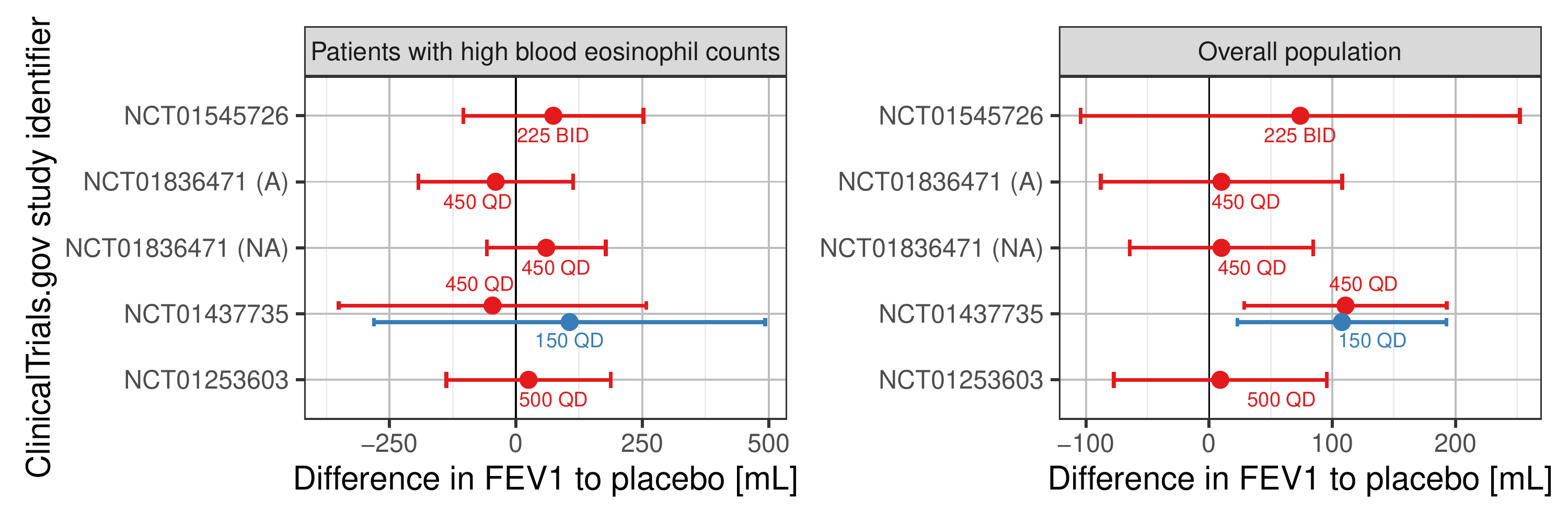}
\end{figure}

Fevipiprant was studied in four Phase 2 RCTs in asthma and the results of these studies for the FEV\textsubscript{1} endpoint are summarised in Figure~\ref{fig:fev1rcts}.

\begin{figure}[htb!]
 \caption{Point estimate and 95\% confidence interval, and predictive distribution with median and 95\% prediction interval for the mean in a new study}
\includegraphics[width=0.98\textwidth]{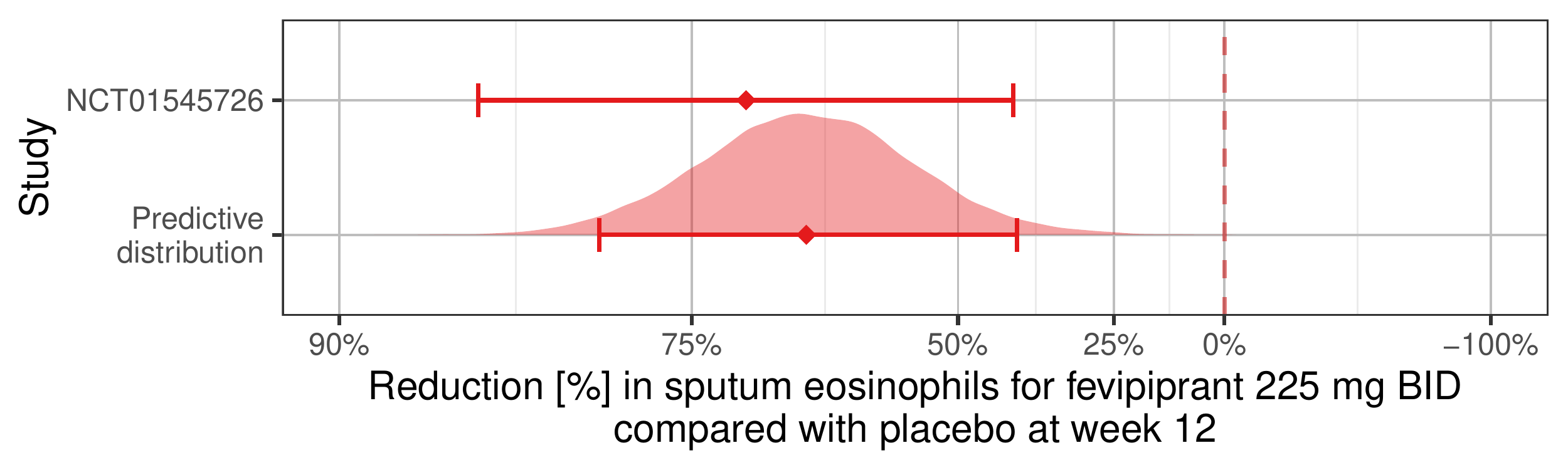} \label{fig:sputum}
\end{figure}

\begin{enumerate}
    \item A Proof of Concept RCT (ClinicalTrials.gov identifier NCT01253603) with a 4 week treatment duration in patients on reliever therapy did not show an effect of fevipiprant on the primary endpoint of FEV\textsubscript{1} in the overall trial population, but more favourable results were seen for a subgroup of more severe patients~\citep{Erpenbeck_2016}. 
    \item A dose finding RCT (NCT01437735) with a 12 week treatment duration~\citep{Bateman_2017} was the basis of the selection of one of the Phase 3 doses. 
    \item A 12-week RCT looked at potential differences in effects in patients with atopic and non-atopic asthma (NCT01836471). 
    \item Finally, there was a RCT (NCT01545726) that showed a reduction of sputum eosinophil counts after 12 weeks of treatment with fevipiprant compared with a placebo group~\citep{Gonem_2016}. The ratio of a 3.5-fold (95\% CI 1.7 to 7.0; p=0.0014) lower ratio of geometric means in sputum eosinophil counts from baseline to the end of treatment compared with placebo in this study was a key rationale for initiating Phase 3 trials investigating an effect on asthma exacerbations~\citep{Brightling2019}. Figure~\ref{fig:sputum} shows these trial results, as well as a predictive distribution for the true value of this ratio in a new study given the results of this study. The prior used for the predictive distribution was based on an industry benchmark as described in Section~\ref{subsec:PoSframework}~\citep{hampson2021}.
\end{enumerate}

\begin{figure}
 \caption{Effects of anti-inflammatory asthma therapies on sputum eosinophil counts and exacerbation rates compared with placebo: Estimates with 95\% confidence intervals for exacerbation rate ratios and ratio of geometric mean (vs. placebo) ratios of sputum eosinophil levels at the end of the study compared with baseline (Panel A), and meta-regression using random drug effects on intercept and slope of relationship, as well as random study effects (Panel B); Studies 10 and 11 are the two parts of study NCT02414854 that were not blinded against each other.}
\includegraphics[width=0.98\textwidth]{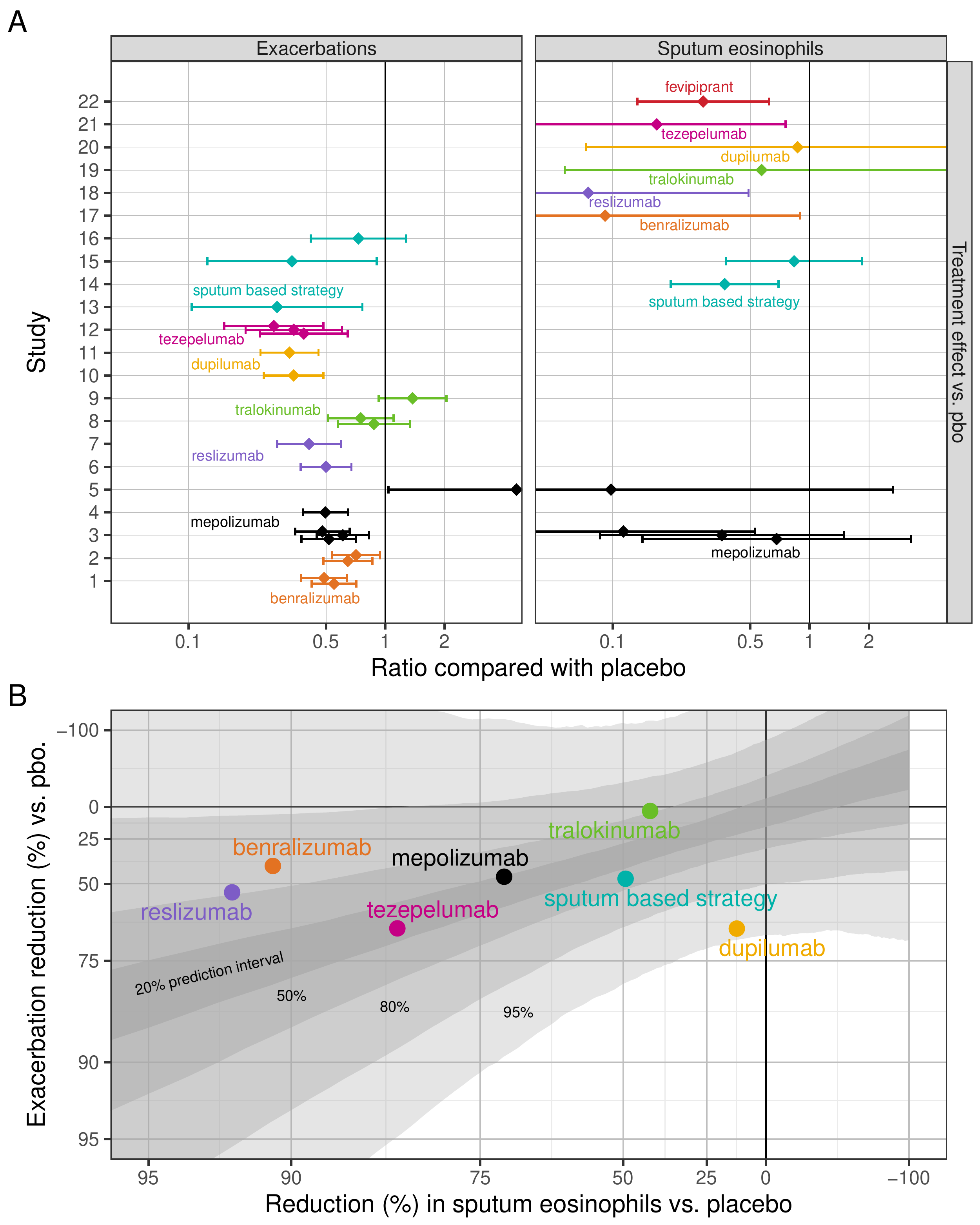}  \label{fig:sputumvsexb}
\end{figure}

A number of anti-inflammatory treatments that lower sputum eosinophil counts have been shown to reduce exacerbation rates in asthma patients with elevated sputum eosinophil counts~\citep{Petsky_2010}. This evidence was mostly generated with corticosteroids, but suggests that sputum eosinophil counts may be a surrogate for a reduction in exacerbations. As part of the evidence dossier for this expert elicitation, we assembled more recent evidence from 22 trials of other drug classes~\citep{Green_2002, Chlumsk__2006, Jayaram_2006, Nair_2009, Fleming_2011, Castro_2011, Pavord_2012, Laviolette_2013, Wenzel_2013, Ortega_2014, Gauvreau_2014, Castro_2015, Bleecker_2016, FitzGerald_2016, Corren_2017, Panettieri_2018, Russell_2018, Castro_2018}. The data are shown in Panel A of Figure~\ref{fig:sputumvsexb} and the results from a Bayesian meta-regression model are shown in Panel B of the figure. Without data from a variety of different drugs, this meta-regression would be highly questionable, because then its findings might only apply to a specific mode of action. Note that some of these data were not available at the time the Phase 3 program for fevipiprant was started. 

For the question of the likely effect of fevipiprant on FEV\textsubscript{1} in asthma patients with blood eosinophil counts $\geq 250$ cells/$\mu$L, the evidence dossier presented the Phase 2 results for the overall population, as well as for subgroups defined by blood eosinophil counts (see Figure~\ref{fig:fev1rcts}).

In addition, the evidence dossier gave details of the fevipiprant Phase 3 program, and discussed the strengths and limitations of the available evidence that the experts needed to bear in mind.

\subsection{Choice of quantities of interest for elicitation}\label{subsec:qoi_choice}

The QoI to be elicited were chosen based on their importance for meeting the success definition of the PoS framework and lack of evidence to directly inform a predictive distribution. The global project team considered the results in the two exacerbation trials (NCT02555683 and NCT02563067) in the pre-specified subgroup of patients with high eosinophil counts to be the most important to fulfil the TPP. These 1-year exacerbation RCTs compared two doses of fevipiprant with a placebo on top of continued standard of care therapy in severe asthma patients. The rate of asthma exacerbations (TPP target: $\geq$ 40\% relative rate reduction compared with placebo) was the primary endpoint of these studies, while the key secondary endpoint of FEV\textsubscript{1} (TPP target: $\geq$ 120 mL improvement in FEV\textsubscript{1} compared with placebo) was considered to be especially important for regulatory approval. There was considerable historical data on the placebo exacerbation rate, the between patient heterogeneity in the exacerbation rate~\citep{Holzhauer_2017} and the variability in FEV\textsubscript{1} so that these quantities did not require elicitation.

The biggest source of uncertainty regarding the PoS was about the effects of fevipiprant on asthma exacerbations and FEV\textsubscript{1}, as well as about their correlation. For this reason, these were identified as the QoIs for the expert elicitation.
We carefully chose the phrasing of the questions about the QoIs to make it easy for the experts to think about them and express their judgements. 

We decided to use the extension method to elicit judgements about the relative rate reduction in exacerbations conditional on a specified reduction in sputum eosinophils, and to use the copula method to elicit the association between the two QoIs. On that basis, we formally defined the following three QoIs:
\begin{itemize}
    \item $X$ is the average reduction in moderate to severe asthma exacerbations achieved by fevipiprant compared to placebo over the population of eligible patients,
    \item $Y$ is the average reduction in sputum eosinophil counts achieved by fevipiprant compared to placebo over the population of eligible patients,
    \item $Z$ is the average increase in FEV\textsubscript{1} achieved by fevipiprant compared to placebo over the population of eligible patients.
\end{itemize}
Eligible patients are defined as matching the inclusion criteria for the NCT02555683 and NCT02563067 Phase 3 trials and having blood eosinophil counts of at least 250 cells/$\mu$L. Note that because we had already derived the marginal predictive distribution in Figure~\ref{fig:sputum} for the reduction $Y$ in sputum eosinophil counts from Phase 2 data, the extension method for the QoI $X$ required only conditional distributions to be elicited.

The choice and phrasing of the QoIs in elicitation is an important early task.  Quantities must be clearly and unambiguously defined, in terms that are familiar to the experts. It must be clear that each quantity has a unique, well-defined (but unknown) value.  We chose to elicit treatment effects compared with placebo as percentage reductions in exacerbations and improvements in FEV\textsubscript{1}, because these are widely used effect measures in asthma trials commonly expressed in these terms that were familiar to the experts.  The effects are defined as averages over all potential patients so that they have  well-defined and unique values. The experts would be asked for their judgements on questions such as: 
\begin{enumerate}
	\item Given that an anti-inflammatory drug reduces sputum eosinophil counts by $Y$, what do you judge to be the likely values for the relative exacerbation rate reduction $X$ in eligible patients?
    \item What do you judge to be the likely values for the difference $Z$ between fevipiprant and placebo in FEV\textsubscript{1} in millilitres (mL) in eligible patients?
   \item Given the judgements about the reduction in exacerbations and the change in FEV\textsubscript{1} caused by fevipiprant, how likely do you judge it to be that both $Y$ and $Z$ will be on the same side of your median values?
\end{enumerate}

\subsection{The elicitation workshop}

\subsubsection{Experts for the elicitation workshop}

In order to capture the full range of opinions and differing past experiences amongst experts, a group of company internal experts was convened.  The 5 selected experts all had extensive experience in drug development in the respiratory area. Two were part of the fevipirant team (a clinician and a statistician), while 3 were not members of the fevipiprant team (a clinician, a translational medicine expert and a regulatory affairs expert). These experts were selected, because the QoIs appeared to be related to clinical trials and understanding mechanistic considerations around the drug efficacy. We wanted at least some of this key expertise to be from outside of the fevipiprant project team to ensure an outside opinion would be heard. A statistician was considered important to provide a perspective on the available evidence and the expert in regulatory affairs was selected due to a broad experience with multiple previous programs.

Prior to the elicitation workshop, all experts were encouraged to work through an online course on expert elicitation~\citep{shelf_e_learning_course} and they were guided through a practice exercise by the facilitator at the start of the workshop.

\subsubsection{Conduct of the elicitation workshop}

The elicitation workshop was an in-person 4-hour meeting with one facilitator, one recorder and five experts. While the facilitator guided the meeting and asked the experts questions, the role of the recorder was to operate the SHELF software, project relevant visualisations for the experts and to take minutes of the meeting.

\subsubsection{Elicitation of first quantity of interest}

The median of the marginal distribution of $Y$ shown in Figure~\ref{fig:sputum} --- based on a Bayesian analysis of Phase 2 sputum eosinophil data --- was a 66\% reduction (80\% interval from 52 to 76\%). Round numbers are easier for experts to condition on, and so, for the first QoI, the median of 66\,\% was rounded to 65\,\%. Thus, the experts were first asked for their judgement on $Y$ conditional on $X$ being a 65\% reduction in sputum eosinophil counts. 

For the individual judgements about this QoI, the tertile method was used. Each expert first independently wrote down their plausible range for the QoI, followed by their median and the points that divide the plausible range into equally probable thirds. At each step the experts were asked to challenge their own judgements. For instance, after specifying their plausible range, experts were asked to consider their reaction if a large study estimated $X$ to be outside that range; would they acknowledge that their range was too narrow, or would they be suspicious of the reported estimate?  If their reaction would be the former one, then they should widen their plausible range.

\begin{figure}
 \caption{Distributions elicited from individual experts, linear pool of these distributions and group judgements}
\includegraphics[width=0.98\textwidth]{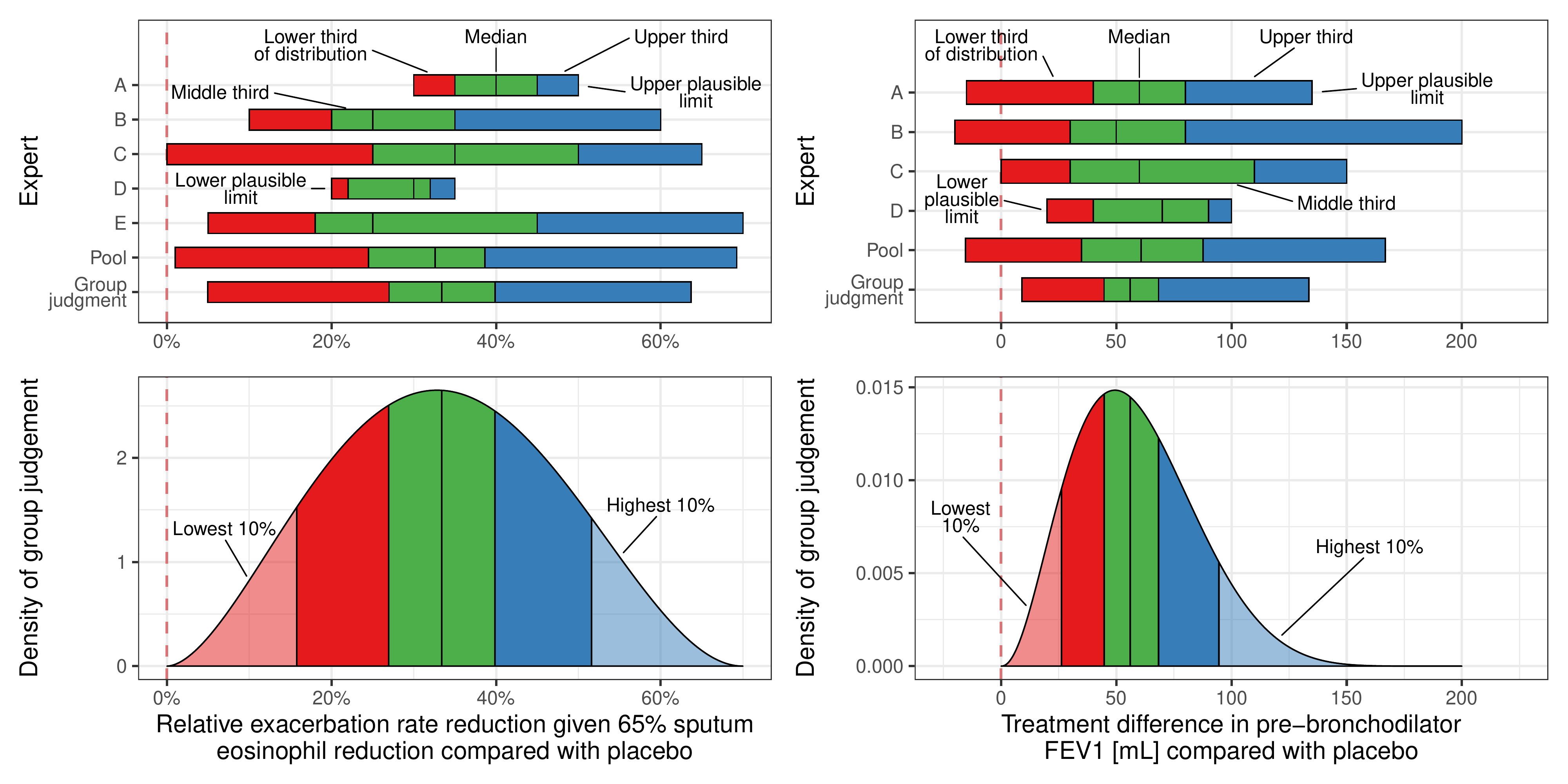} \label{fig:elicited1}
\end{figure}

Then the individual judgements were revealed to the group and the experts were asked to explain their judgements. In this wide-ranging discussion, a number of points were raised and the main arguments were recorded using the SHELF templates. Afterwards, consensus judgements were obtained using the probability method: experts were asked what probability RIO (the Rational Impartial Observer) would assign to the relative exacerbation rate reduction being less than 25\%, greater than 40\% and less than 35\%. After significant discussion, the group agreed that RIO would assign probabilities of 30\%, 30\% and 50\%, respectively. A Beta(2.81, 3.05) distribution scaled to a plausible range of 0 to 70\% was fitted to these judgements and shown to the experts. The experts felt that this distribution, with a median at 33.4\% (90\% credible interval 11.9 to 55.8\%), adequately represented their knowledge.
The result of this elicitation was a distribution for $X$ (exacerbation reduction), given that $Y$ (sputum eosinophil reduction) is 65\%. The results of the individual judgements and the group judgement are shown on the left-hand side of Figure~\ref{fig:elicited1}.

\begin{figure}
 \caption{Piecewise-linear median model for the elicited medians (Panel A) and conditional distributions for the relative exacerbation rate reduction across the range of plausible effects on sputum eosinophil counts (Panel B)}
\includegraphics[width=0.98\textwidth]{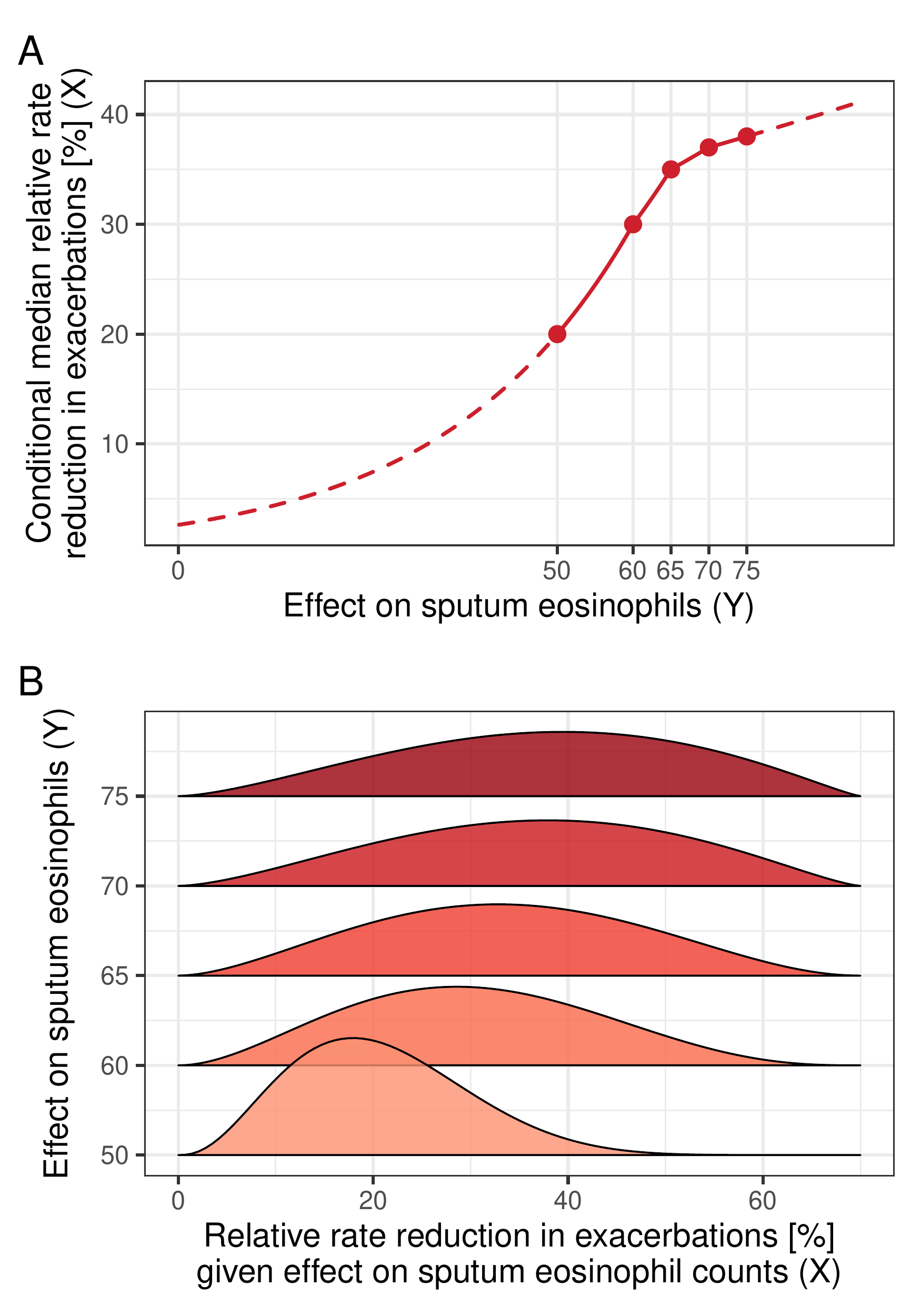} \label{fig:condelicited}
\end{figure}

Then, the experts were asked for their conditional judgement about the median percentage reduction in exacerbations given an effect on sputum eosinophil of 50\%, then for 75\%, 60\% and 70\%. These numbers correspond approximately to 10\%, 90\%, 25\% and 75\% points of the marginal predictive distribution for effects of fevipiprant on sputum eosinophil counts, respectively. Thus, they characterise conditional judgements across the bulk of this distribution. Their order was chosen in order to minimise known sources of cognitive bias and to ensure that experts needed to think carefully about each judgement. The elicited medians are shown in Panel A of Figure~\ref{fig:condelicited}.

It was agreed that over the plausible range of effects on sputum eosinophil counts, there was no probability that the drug could increase the number of exacerbations, because the assumption that fevipiprant reduced sputum eosinophils indicated at least some positive benefit. It was therefore appropriate to model the distributions of exacerbation reductions at intermediate sputum eosinophil effects through a log transformation --- i.e.\@ to assume that $\text{median}(\log(X|Y))$ is a piecewise linear function of Y. The experts were shown the resulting median relationship shown in Panel A of Figure~\ref{fig:condelicited} and agreed that it represented a reasonable RIO opinion. 

Using the log transformation, the conditional distribution given $Y = 65\%$ was assumed for $X$ conditional on other values of $Y$, but scaled to follow the elicited median model --- i.e.\@ we shifted the median of each Beta-distribution according to Panel A of the figure and kept the variance on the log-scale constant. The recorder showed the experts the resulting conditional distribution plot in Panel B of Figure~\ref{fig:condelicited}. The facilitator pointed out how the scaling had resulted in less uncertainty conditional on $Y = 50\%$ but more conditional on $Y = 75\%$.  The experts confirmed that this was a reasonable representation of their beliefs.

The elicitation of the first QoI was now complete and the required (marginal) distribution for $X$ was computed by Monte Carlo simulation by combining the elicited conditional relationship with the predictive distribution for $Y$ from Figure~\ref{fig:sputum}. It is shown in the top-most panel of Figure~\ref{fig:resvsprior}.

\subsubsection{Elicitation of further quantities of interest}

The elicitation for the second QoI then proceeded using the tertile method for individual judgements, followed by a discussion and, again, using the probability method for the consensus judgement. The resulting judgements are shown on the right-hand side of Figure~\ref{fig:elicited1}.

\begin{figure}
 \caption{Point density plot of elicited joint distribution for treatment effect on exacerbations and FEV\textsubscript{1} based on 10,000 Monte Carlo samples}
\includegraphics[width=0.98\textwidth]{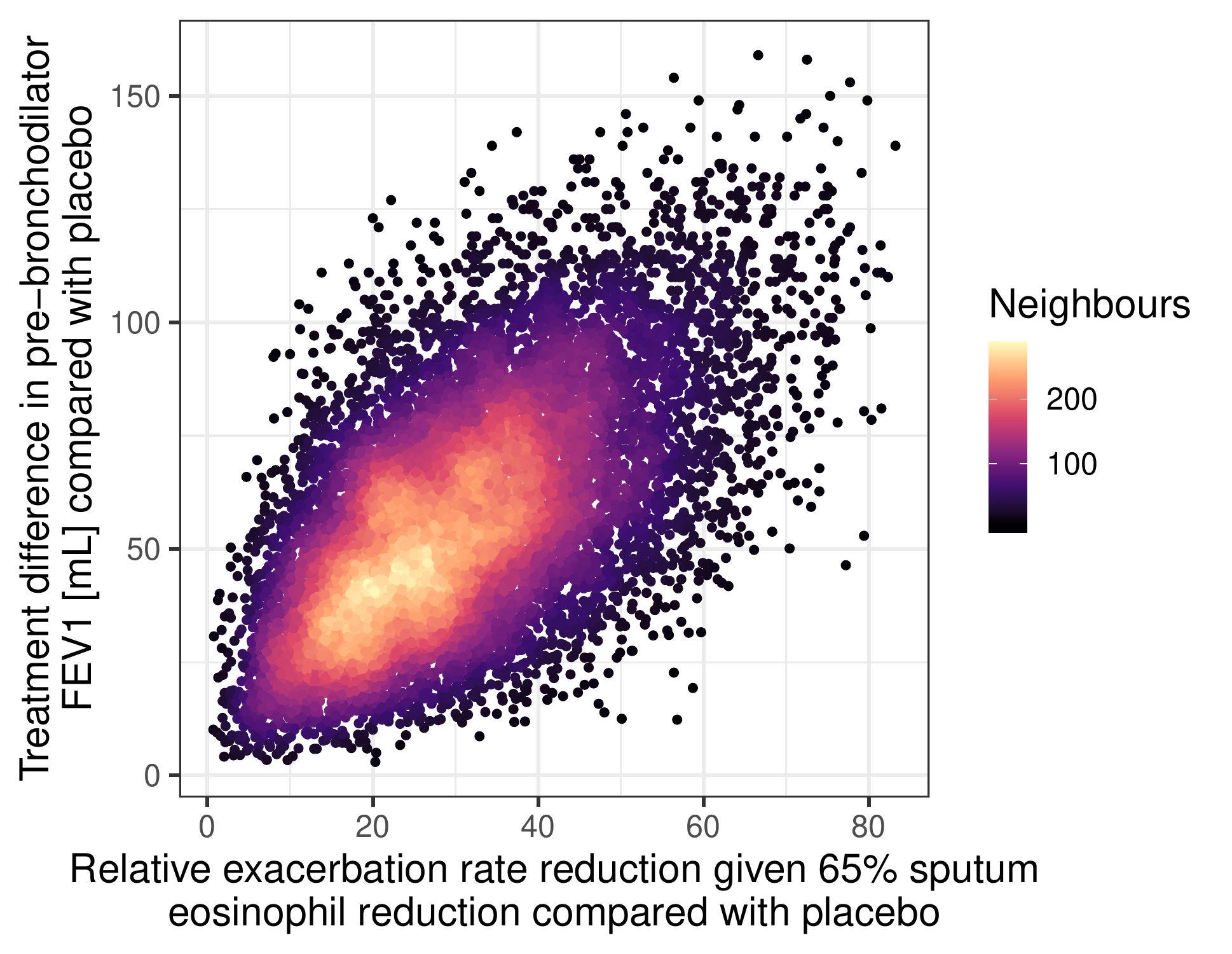} \label{fig:joint}
\end{figure}

The joint distribution of the treatment effects on exacerbations and FEV\textsubscript{1}, $X$ and $Y$, was then elicited using the copula method.  The correlation was elicited through the concordance probability, i.e.\ RIO's judgement of the probability that the true values of $X$ and $Y$ would both be on the same side of their elicited medians. The experts found the concordance probability difficult to judge. After the facilitator gave an alternative explanation in terms of the conditional probability that one variable was above its median given that the other was above its median, a concordance probability of 0.7 was tentatively agreed by the experts. The experts were shown a graphic similar to Figure~\ref{fig:joint} for the case of a concordance probability of 0.7 and found it very helpful and in accord with their expectations. Alternative concordance probabilities were explored using the same graphical display. The correlation was too tight with 0.8 concordance and the experts felt that there was appreciable positive correlation so 0.5 concordance was not considered appropriate. The resulting joint distribution is shown in Figure~\ref{fig:joint}.

\subsection{Probability of success calculation}
\label{subsec:PoSframework} 

We already described the basic aims of the newly introduced PoS framework at Novartis at a high level in Section~\ref{sec:motivate}. Its practical application involves the following four steps~\citep{hampson2021}. First, a benchmark probability of approval for a project at the start of Phase 2 is estimated based on a small number of program characteristics by a logistic regression model trained on a database of drug development projects. Second, a Bayesian analysis is conducted, in which the prior for the efficacy effects is set based on the benchmark probability of efficacy success in both Phase 2 and 3. This prior is then used in combination with Phase 2 data to obtain a posterior distribution for drug efficacy. Phase 3 studies are then simulated using samples from the posterior in order to estimate the probability of the key efficacy endpoints meeting TPP criteria in the Phase 3 program. Benchmark information is also used to account for the risk of program failure due to an unexpected safety issue and of not obtaining regulatory approval despite a successful Phase 3 program. Third, a program risk assessment is done to capture other risks not already covered by the previous calculations. This assessment is then used to adjust the probability of a registration with a label meeting TPP criteria to obtain the PoS. The adjustment in this step was also determined using elicitation process. Finally, in exceptional circumstances a fourth step allows for an adjustment for factors not captured by the preceding three steps.

In this case study, the Bayesian analysis in the second step of the PoS approach could not directly inform the PoS of the Phase 3 program due to the differences in endpoints and population between Phase 2 and 3. Thus, the results of the Bayesian analysis for sputum eosinophil counts in Figure~\ref{fig:sputum} were linked to the efficacy on asthma exacerbations in Phase 3 using an expert elicitation in the manner described in Section~\ref{subsec:shelf extension}. In contrast, the effect of fevipiprant on FEV\textsubscript{1} was elicited directly from the experts and the joint distribution of the efficacy of fevipiprant for both endpoints was then obtained as described in Section~\ref{subsec:shelf copula}.

For pragmatic reasons the Novartis PoS approach foresees that only one or two key endpoints should be considered in the definition of success. For this reason, it was decided to ignore the other two key secondary endpoints (asthma control questionnaire and asthma related quality of life questionnaire) of these Phase 3 trials for the purposes of the PoS calculation.

\subsubsection{Calculation of PoS estimates}

The estimated benchmarks for the first indication of a respiratory orally administered small molecule without a FDA breakthrough designation were:
\begin{itemize}
\item a Phase 2 success probability of 24\%,
\item a Phase 3 success probability of 60\% conditional on Phase 2 success, and
\item an approval probability of 94\% conditional on Phase 2 and 3 success.
\end{itemize}

The program risk assessment~\citep{hampson2021} considered the majority of categories to fall into the lowest risk category with one question falling into the intermediate risk category.

When these numbers were combined with simulated Phase 3 outcomes based on the elicited quantities, a PoS of 4\% was calculated. The main hurdle was FEV\textsubscript{1} and the high TPP target for exacerbation reduction. If one only considered a TPP requiring a relative exacerbation reduction of 30\% with no requirements for FEV\textsubscript{1}, the PoS became 41\%.

\subsection{Timelines}
The whole PoS process required approximately 2 months. After an initial review, we identified that an expert elicitation workshop would be needed. On 28 May 2019, we identified the facilitator for the workshop and compiled a list of candidate dates. In the meantime, the team worked to assemble an evidence dossier. By 12 June, we had arranged a elicitation workshop on 12 July after confirming the availability of five experts. By 1 July, the evidence dossier had been drafted by the biostatistics team, was shared with the facilitator and recorder, and was finalised on 8 July after a review by internal experts, four days before the workshop. One learning was that we should have shared the dossier with the experts earlier in order to allow them to provide feedback on its contents so that additional evidence could have been introduced up-front. On 12 July the workshop took place using version 4 of the SHELF methodology and on 20 July 2019 the final report of the elicitation meeting was issued. All recordings from the meeting were made using the templates provided as part of the SHELF documents package and participants were kept anonymous in these minutes by using the letters A to E for the experts, as well as Z for the facilitator.

\subsection{Phase 3 results}

\begin{figure}
 \caption{Implied distribution for true effect of fevipiprant 450 mg QD on exacerbations and  FEV\textsubscript{1} based on elicited expert judgements, and study results in the high blood eosinophil subgroup of the Phase 3 exacerbation trials}
\includegraphics[width=0.98\textwidth]{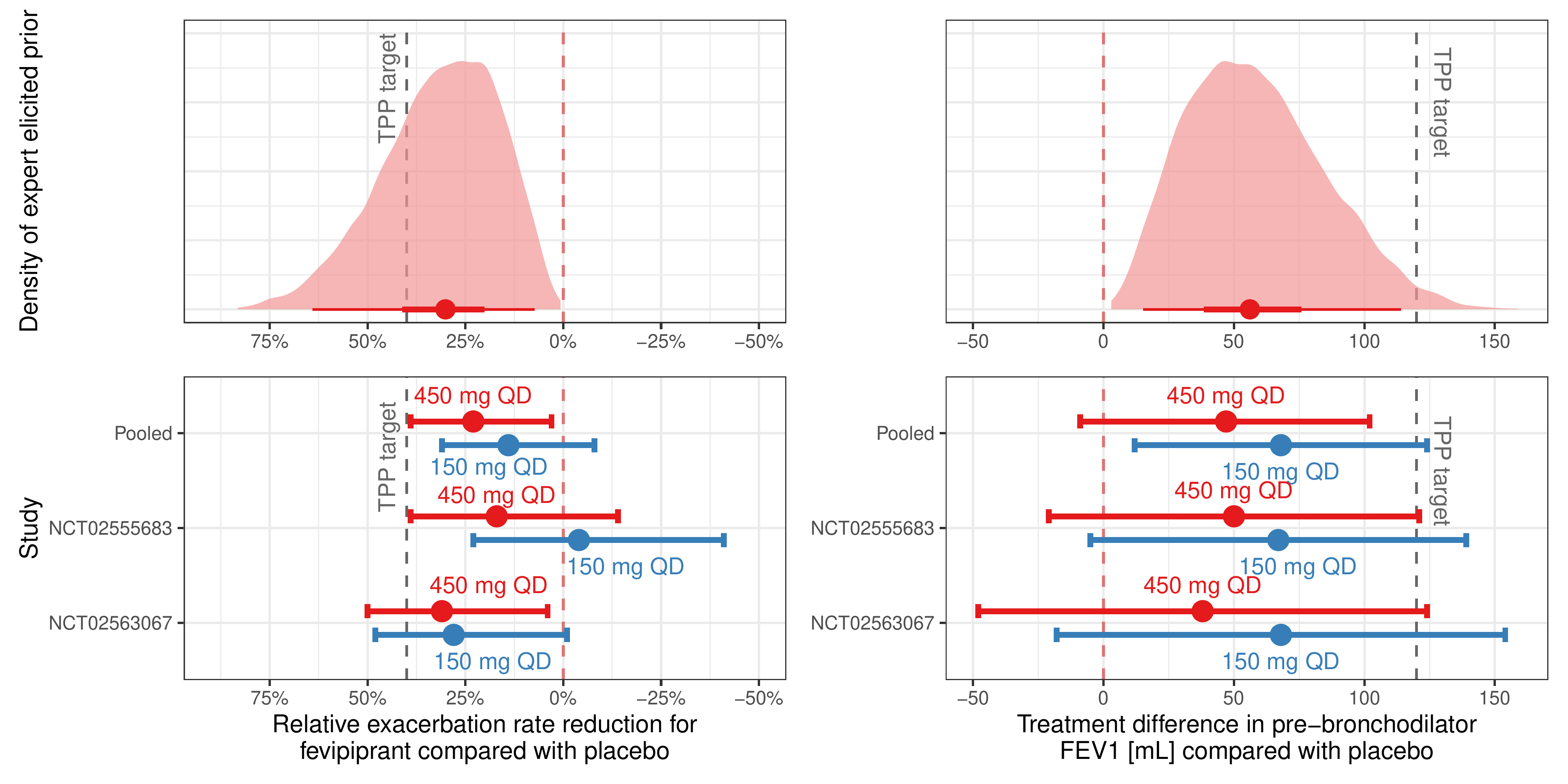} \label{fig:resvsprior}
\end{figure}

The results of the Phase 3 trials, for which we conducted the expert elicitation, are shown in Figure~\ref{fig:resvsprior}. As can be seen only one comparison within one of the two trials was associated with a confidence interval that excluded no effect, but this result was not considered statistically significant after an adjustment for multiplicity~\citep{Brightling_2020}. The results of the Phase 3 trials are very informative in the sense that the 95\% confidence intervals essentially exclude the TPP targets.

These results are consistent with the elicited prior information from the experts: the experts essentially excluded the possibility that the true effect of the studied fevipiprant doses on FEV\textsubscript{1} meet the TPP target, while for the primary exacerbation endpoint, the experts judged that there was a reasonable possibility that the true effect was at or above the TPP target. On the basis of these Phase 3 results Novartis did not pursue a filing for an indication in asthma. 

\section{Discussion}\label{sec:discussion}

The quality of decisions in the presence of uncertainty can be improved by taking the judgements of experts based on the available evidence into account. When stakes are high, as with major investment decisions by a pharmaceutical company, the necessary effort and cost of obtaining experts' judgements is negligible compared to the cost of a wrong decision. This is one of the reasons why the new Novartis PoS framework, which is applied for the decision to initiate pivotal trials for a project, recommends expert elicitation when substantial direct evidence about QoIs is not available. The SHELF extension method and the SHELF copula method address two common scenarios in this setting: when we extrapolate the evidence from surrogate endpoints to Phase 3 endpoints, and when how much a drug affects one endpoint changes how much we judge it to affect other endpoints.

There are currently no published examples of how to apply these methods as part of the SHELF protocol in the pharmaceutical industry. Therefore, we felt it would be helpful to share an example illustrating the full extent of real-world complexities and the relevant practical considerations. This will hopefully help others that wish to use expert elicitation to inform clinical drug development or other types of high stakes decisions.

We do not wish to overemphasise the outcomes from a single example. Nevertheless, the close alignment between the experts' group judgements with the trial outcomes, which were not known to the experts at the time of the elicitation workshop, supports the validity of expert elicitation in drug development. If a similar elicitation outcome had been available at the time of the decision to start the Phase 3 program for fevipiprant, it would have suggested a lower PoS than assigned at the time and may have led to re-evaluation of the assumptions regarding the secondary FEV\textsubscript{1} endpoint. However, this proof of concept for elicitation as part of a new PoS framework was performed 4 years after this decision and used information that only became available subsequently.

The project team noted that the evidence dossier and the discussions in the elicitation workshop were extremely helpful for assembling and understanding the existing evidence on the efficacy of the drug. It may sometimes be the case that teams are very well aware of the clinical trials conducted for their product, but have not systematically reviewed the indirect evidence that is available from other sources. After the elicitation workshop the experts expressed that they appreciated the structured and scientific process, that they found the methodology intuitive, and that they were positively surprised how fully non-statisticians could participate in the workshop.

While we describe a particular example of an elicitation workshop, we have now run several similar workshops at Novartis and some of the authors of this paper have several years of experience of doing so with other clients. On this basis, we offer a number of practical recommendations. It is important to start preparing the evidence dossier as early as possible so that experts and other stakeholders can give feedback prior to a  workshop. This is also an opportunity to let senior leaders with strong positive opinions on projects provide the evidence they wish to be considered. Additionally, it can be difficult for experts to free their agenda for long workshops and we have found that people find it hard to concentrate in virtual meetings for as long as in in-person workshops. This has led us to investigate options for eliciting individual judgements prior to the main workshop. It is also important to clearly communicate how elicitation results will be used. In the context of the PoS of drug development programs, this meant making it clear that the resulting probability is not the sole determinant of funding for a project. We now routinely remind teams that investment decisions will also be based on other factors such as the costs of development, market opportunity and unmet medical need.

\section{Acknowledgments}
We thank Ana-Maria Tanase, Christian Hasenfratz and Hanns-Christian Tillmann for being experts for the asthma case study, as well as Kelvin Stott, Giovanni Della Cioppa and Karine Baudou for their support of the pilot phase of the Novartis PoS initiative.

\clearpage
\bibliographystyle{plainnat}
\bibliography{bibliography}

\end{document}